 \documentclass[a4paper, floatfix]{revtex4}
\usepackage{vmargin,xcolor} 
\usepackage[english]{babel} 
\usepackage[utf8]{inputenc} 
\usepackage{graphicx,float,setspace,amssymb,amsmath} 

\newcommand{\equa}[1]{\begin{eqnarray} \label{#1}} 
\newcommand{\auqe}{\end{eqnarray}} 
\newcommand{\tab}[1]{\begin{tabular}{#1}} 
\newcommand{\bat}{\end{tabular} \\ } 
\newcommand{\blanc}{\makebox[1.1 cm]{ }}

 \newcommand{\be}{\beta} 
 \newcommand{\ep}{\epsilon} 
 \newcommand{\gD}{\Delta}  
\newcommand{\s}{\sigma} 
 
\newcommand{\dd}{\hat{m}_i\hat{m}_j -3 (\hat{m}_i \hat{r}_{ij})(\hat{m}_j \hat{r}_{ij})}

\providecommand{\abs}[1]{\left\vert#1\right\vert} 
 
\begin{document} 
\selectlanguage{english} 
\title 
{{\bf Size and polydispersity effect on the magnetization of densely packed magnetic nanoparticles. 
}} 
\author 
{ Vincent Russier~$^{a)}$, Caroline de Montferrand~$^{b)}$, Yoann Lalatonne~$^{b)}$ and Laurence Motte~$^{b)}$}
\affiliation {
~$^{a)}$ ICMPE, UMR 7182 CNRS and UPEC, 2-8 rue Henri Dunant 94320 Thiais, France, \\
~$^{b)}$ CSPBAT UMR 7244 CNRS and University Paris 13, 93017 Bobigny, France. 
} 
%
\thispagestyle{empty} 
\begin{abstract} 
The magnetic properties of densely packed magnetic nanoparticles (MNP) assemblies are
investigated from Monte Carlo simulations. The case of iron oxide nanoparticles
is considered as a typical example of MNP. The main focus is put on particle size 
and size polydispersity influences on the magnetization curve. 
The particles are  modeled as uniformly magnetized spheres isolated one from each 
other by a non magnetic layer representing the organic coating.
A comparison with recent experimental results on
$\gamma-$Fe$_2$O$_3$ powder samples differing by their size is given.
\end{abstract}
\maketitle 
\eject 
\section {Introduction} 
\label{intro} 
The physics and chemistry of nanoscale magnetic particles (MNP) still gives 
rise to an important research activity
due both to their wide range of potential applications and their own fundamental
interest \cite{dormann_1997, skomski_2003, majetich_2006, bedanta_2009}. Among the large variety of 
MNP, iron oxide based ones $\gamma-$Fe$_2$O$_3$ and Fe$_3$O$_4$ take a particular place
in the field of biological and medical applications because of their bio-compatibility and 
suitable superparamagnetic properties. 
To translate intrinsic properties of nanoparticles to various applications, there is a need 
to control nanoparticle dispersions. Consequently nanoparticles are usually coated by an 
organic surfactant \cite{lalatonne_2005, lalatonne_2008}
in order to prevent aggregation. The influence of this non magnetic layer and then the 
nanoparticles contact distance play a major role on collective magnetic properties 
\cite{ lalatonne_2004}.
A complete understanding of the macroscopic magnetic 
properties of MNP assemblies in terms of their individual intrinsic characteristics on the 
one hand and of the size distribution and volume concentration on the other hand is of crucial 
importance. Indeed this is a mean to get informations on the relevant parameters of the distribution 
and MNP properties from the magnetic measurements. Two key features which strongly influence 
the macroscopic magnetic properties of these systems are the magnetic structure at the 
particle scale, where core shell structure and spin canting effect can be invoked 
\cite{coey_1971, tronc_2000, tronc_2003, battle_2002}
and the size distribution generally described through a lognormal law for the diameters distribution.

At temperatures higher than the blocking temperature T$_b$ where the MNP are in the 
superparamagnetic regime \cite{dormann_1997, bedanta_2009} and in case of weak interparticle 
interactions, namely for both small particles concentrations and in the absence of cluster 
formation the physical properties of MNP assemblies are well understood.
The magnetization curve, $M(H)$ of the whole assembly follows then a Langevin like function
weighted by the diameter distribution function and eventually modified in order to
take into account a core-shell structure
\cite{morales_1999, fortin_2007, chen_2009}. 
Moreover the one-body magnetocrystaline anisotropy 
energy of the MNP can also be taken into account and this modifies the $M(H)$ curve from the
original Langevin function \cite{garcia-palacios_2000, wiekhorst_2003}. 
The core shell structure of the MNP may consist simply of the inclusion of a magnetic 
dead layer at the surface of the MNP \cite{battle_2002, tronc_2003} or of the introduction of
an additional paramagnetic component in the MNP \cite{chen_2009}. The symmetry breaking at the 
surface can lead to surface effects on the anisotropy energy of each MNP with noticeable effects 
on the $M(H)$ curve \cite{yanes_2007, margaris_2012, kachkachi_2006} .  
In case of diluted assemblies of spherical MNP when the particles are non or weakly 
interacting, the non interacting particles type of approach of the magnetization curve leads to
a reasonable determination of the characteristics of the individual particles 
and of the size distribution namely the median diameter $d_m$ and the $\ln(d)$
standard deviation $\s$. However, when the NP concentration increases, the interparticles
interactions must be taken into account.
These ones which for spherical and well coated MNP include 
mainly the interactions between the MNP magnetic dipoles (DDI), have been widely
studied and a large amount of works and methods are thus available going from
mean field approximation, thermodynamic perturbation theory (TPT) \cite{jonsson_2001, margaris_2012} 
for weakly interacting systems to numerical simulations for moderate to strongly interacting
systems \cite{kechrakos_2000, chantrell_2000, margaris_2012}.
The mean field and TPT provide an illustrative physical picture of the relation
between the local structure and either the magnetization in terms of the applied field
or the susceptibility. For instance the demagnetizing field effect depending 
on the external shape of the system, is well reproduced by the TPT \cite{jonsson_2001}. 
As a link between TPT and numerical simulations, the description based on the interaction fields
distributions \cite{al-saei_2011} which explains the DDI induced reduction of the magnetization
of an isotropic system as a generalization of a similar result obtained using the TPT and 
suggests that the DDI induced reduction of the magnetization is not related to an 
antiferromagnetic behavior. However, for strongly interacting systems, as in 
lyophilized powder samples or high concentration MNP assemblies embedded in a non 
magnetic matrix the numerical simulations seem more adapted. Although numerical 
simulations of magnetic properties of MNP assemblies are now many, a systematic 
study of the mean size and polydispersity effects especially for randomly organized
particles with high concentration is still missing. 

The aim of this work is to investigate this problem and to interpret 
recent experimental measurements \cite{de-montferrand_2012} on powder samples 
of maghemite MNP assemblies differing by their median size. 
We present a Monte Carlo (MC) simulation of the mean particle size and 
polydispersity effect on the DDI in random and densely packed spherical clusters of 
coated spherical maghemite MNP. Our main purpose is to model the case of lyophilized 
powders or high concentration of particles embedded in a non magnetic matrix. 
A particular attention is paid to the linear susceptibility $\chi$, and its dependence
on the median size of the size distribution. It is found that $\chi$ as a function of 
$d_m$ may present a plateau, leading to a quasi independence of the magnetization 
with respect to $d_m$ in the vicinity of the low external fields. The magneto 
crystalline anisotropy is then shown to play a role for larger values of the field
when the particles remain in the superparamagnetic regime in agreement with the
findings of Ref. \cite{garcia-palacios_2000, garcia-palacios_2000b} for non interacting particles, 
in the TPT regime \cite{jonsson_2001} and in preceding MC simulations
\cite{chantrell_2000, kachkachi_2005, margaris_2012}.
As an application, we focus on the experimental magnetization curves of Ref.\;\cite{de-montferrand_2012}.
\section{Model for densely packed assemblies } 
\label{dense_packed} 
The model we use is designed to simulate the properties of
either lyophilized powders samples or high concentration 
nanoparticles assemblies embedded in non magnetic matrix.
As is usually done to model single domain MNP,
the nanoparticles are modeled as non overlapping spheres bearing at their center a
permanent point dipole representing the uniform magnetization of the particle (super spin).
The moment of each particle is equal to its volume times the bulk magnetization, $M_s$,
which means that neither spin canting effect nor magnetic dead layer at the particle
surface is considered. We also include the magneto crystalline anisotropy 
with the same anisotropy constant $K_1$ on all particles. The particles are supposed to 
be coated by a non magnetic layer of thickness $\gD$/2, representing the usual coating by
organic surfactant molecules.
The layer thickness is taken as $\gD$/2 for convenience (see below). 
The particle diameters, $\{d_i\}$ are distributed according to a log-normal law 
defined by the median diameter $d_m$ and the standard deviation $\s$ of $ln(d)$,
\equa{lgn_1}
f(d) = \frac{1}{d \sqrt{2\pi}\s}\exp \left(-\frac{(\ln(d/d_m))^2}{2\s^2} \right) 
\auqe
$d_m$ and $\s$ are related to the mean diameter and the diameter standard deviation $\s_d$ through
$d_1 = d_m e^{\s^2/2}$ and $\s_d\;=\;d_m\sqrt{(e^{\s^2}-1)e^{\s^2}}$. In the following, we use $d_m$ 
as the unit of length, and thus in reduced unit, the distribution function is totally determined
by the single parameter $\s$ which characterizes the system polydispersity.

We consider mainly spherical clusters, where owing to the global shape isotropy 
the demagnetizing effects vanish, with free boundary conditions. 
This choice of large spherical clusters can be justified on the experimental point of
view since upon drying the NP are likely to aggregate in spherical shaped large clusters
which has been confirmed from simulations \cite{lalatonne_2005}.
Our first purpose is to focus on the contribution of the dipolar interactions (DDI)
to the magnetization curve, especially in the moderate to strong coupling regime
when particles surrounded by their coating layer are at contact. 
The geometrical configuration of two particles of different sizes at contact with 
their coating layer is displayed in figure (\ref{shema_contact}).
Moreover, we consider 
temperatures such that the particles of size $d_m$ are superparamagnetic; as 
we shall see later for polydisperse systems 
due to the presence of large particles in the distribution,
this condition may not be strictly fulfilled.
When taken into account the magnetocrystalline anisotropy is considered in its simplest form, 
namely in the uniaxial symmetry and at lowest order \cite{skomski_2003, bedanta_2009}. 
The total energy thus includes the DDI, 
the one-body anisotropy term and the Zeeman term corresponding to the interaction with 
the external applied field $\vec{H}_a = H_a\hat{h}$.
Let $\{\vec{r}_i\}$, $\{v(i)\}$ , $\{\vec{m}_i\}$ and $\{\vec{n}_i\}$ denote the particles locations, 
volumes, moments and easy axes respectively. The total energy of the cluster reads 
\equa{ener_dip1}
E = \frac{\mu_0}{4\pi} \sum_{i < j} m_i m_j \frac{\hat{m}_i\hat{m}_j -3 (\hat{m}_i \hat{r}_{ij})(\hat{m}_j \hat{r}_{ij})}{r_{ij}^3}
    - K_1 \sum_i v(i)(\hat{n}_i\hat{m}_i)^2
    - \mu_0 H_a \sum_{i} m_i \hat{m}_i \hat{h}
\auqe
where hated letters denote unit vectors, $m_i$ are the moment magnitudes, 
$r_{ij} = \abs{\vec{r}_i - \vec{r}_j}$. It is worth mentioning that the consideration of the 
anisotropy term with a fixed easy axes distribution means that the magnetization relax according 
to a N\'eel process \cite{du-tremolet_2000, coey_2010},
namely the particles are considered fixed while their moment relaxes relative to their 
easy axis.  
In this work only the case of a random 
distribution of easy axes is considered. In the following we use reduced quantities; first the energy 
is written in $k_B T_0$ units, $T_0$ being a suitable temperature ($T_0 = 300K$ in the present work) 
and we introduce a reference diameter, $d_{ref}$. 
The reference diameter, $d_{ref}$ is a length unit independent of the size distribution, useful
for the energy couplings, and can be chosen from a convenient criterion independently of the actual structure
of the MNP assembly.
The reduced total energy is given by
\equa{ener_dip2}
\be_0 E &=& - \ep_K^{(0)} \left( \frac{d_m}{d_{ref}} \right)^3 \sum_i d_i^{*3} (\hat{n}_i\hat{m}_i)^2
            - \ep_d^{(0)} \left(\frac{d_m}{d_{ref}}\right)^3 
            \sum_{i < j} d_i^{*3}d_j^{*3} \frac{\dd}{r_{ij}^{*3}} \nonumber \\
        &-& h \sum_{i} d_i^{*3} \hat{m}_i \hat{h}      \nonumber \\
            \ep_d^{(0)} &=& \frac{\be_0 \mu_0}{4\pi} (\pi/6)^2 M_s^2d_{ref}^3 ~~~~
            \ep_K^{(0)} = \be_0 K_1 v(d_{ref})  \nonumber \\
      h &=& \be_0\mu_0 M_s (\pi/6) d_{m}^3 H_a \equiv \left(\frac{d_m}{d_{ref}}\right)^3 \frac{H_a}{H_{ref}}
\auqe
where $\be_0 = (k_B T_0)^{-1}$ and the stared lengths are in $d_m$ unit.
The dimensionless dipolar coupling constant and anisotropy constant are then 
$\ep_d = (d_m/d_{ref})^3\ep_d^{(0)}$ and $\ep_K = (d_m/d_{ref})^3\ep_K^{(0)}$ respectively; the reference
diameter, $d_{ref}$ can be chosen such that 
$\ep_d(d_m = d_{ref}) \equiv \ep_d^{(0)}$ = 1 ;
the reduced external field $h$ coincides
with the usual Langevin variable 
at temperature $T_0$ for a monodisperse distribution with $d$ = $d_m$
In equation (\ref{ener_dip2}), we also introduce
the reference external field, $H_{ref}$ for convenience. 

Concerning the structure in position, the nanoparticles surrounded by their coating layer $\gD$ form an 
assembly of hard spheres of effective diameters $\{d_i + \gD\}$
(see figure (\ref{shema_contact}))  which are arranged in large densely
packed clusters with either a random or a well ordered structure (simple cubic or face centered cubic
lattice). We build these clusters in two steps. First a large stacking of the coated spheres is made in a 
parallepipedic box with the desired structure, random or well ordered. In the random case, this first
step is made from a sequential random rain plus compression algorithm in such a way to maximize the packing
fraction. Doing this we can get a packing fraction $\varphi$ for the effective spheres corresponding to the 
so-called loose random packing \cite{soppe_1990} ($\varphi \simeq$ 0.60 in the monodisperse case).  
Once this first step is performed, we cut within the global stacking the cluster we want to 
study by imposing both the external shape, either spherical or prismatic, and the number of 
particles $N_p$, with typically $N_p \simeq$ 1000.
The central part of some of the clusters used in the present work corresponding to
different values of the polydispersity, $\s$ is shown in figure (\ref{centre_cluster}) to illustrate the 
structures obtained.
It is important to note that because of the coating layer 
of thickness $\gD$/2  the closest distance of 
approach between particles $i, j$ is shifted from $(d_i + d_j)/2$ to $(d_i + d_j)/2 + \gD$ and 
as a result the sum involved
in equation  (\ref{ener_dip2}) depends on the actual magnetic particles concentration of the cluster through 
the value chosen for $\gD$. 
One can rewrite the DDI sum by using another length scale, namely $(d_m + \gD)$ in order to exhibit a contribution 
independent of $\gD$. Doing this, the total DDI energy reads
\equa{sum_ddi_1}
\ep_d^{(0)} \left( \frac{d_m}{d_{ref}} \right)^3\left(\frac{d_m}{d_m + \gD} \right)^3
        \sum_{i<j} d_i^{*3}d_j^{*3} \frac{\dd}{(r_{ij}/(d_m+\gD))^{3}}
\auqe
We recall that the distribution of reduced diameters, $\{d_i^{*}\}$ depends only on the value of $\s$,
which is conserved through a scaling operation corresponding to a change of $d_m$.  The sum 
of equation  (\ref{sum_ddi_1}) is a geometric sum characteristic of the DDI expected, at least for 
small values of $\s$, to be independent of $\gD$ and thus to characterize the reduced DDI sum 
of the most concentrated cluster ($\gD$ = 0) of the structure (s.c., f.c.c., random) considered.
In other words, equation (\ref{sum_ddi_1}) allows to explicit the dependence of the dipolar 
coupling with respect to the particles volume fraction, $\varphi_v$. 
For this we note that $(d_m/(d_m + \gD))^3$ can be rewritten as $\varphi_v/\varphi_m$
where $\varphi_m$ ($\equiv \varphi_v(\gD=0)$) is the maximum value of $\varphi_v$ for the given configuration,
namely the volumic fraction corresponding to the spheres of diameters $\{d_i + \gD\}$.
$\varphi_m$ = $\pi/6$, 0.74, and $\sim$ 0.60 for the simple cubic, fcc and the loose random packed
structures respectively.
Then from (\ref{sum_ddi_1}),
we can introduce an effective dipolar coupling constant, say 
$\ep_d^{(eff)} = \ep_d^{(0)}(d_m/d_{ref})^3(d_m/(d_m + \gD))^3$ which is rewritten as
$\ep_d^{(eff)} = (\varphi_v/\varphi_m)(d_m/d_{ref})^3$,
since $\ep_d^{(0)}$=1.0. 
Now, one can replace both $d_m$ and $\gD$ by say $d_{m2}$ and 
$\gD_2$ respectively in such a way that the total DDI energy remains constant by imposing
\equa{sum_ddi_2}
\left( \frac{d_m}{d_{ref}} \right)\left(\frac{d_m}{d_m + \gD} \right) = 
\left( \frac{d_{m2}}{d_{ref}} \right)\left(\frac{d_{m2}}{d_{m2} + \gD_2} \right)
\auqe
leading to
\equa{d2_delta2}
d_{m2} = \frac{d_m}{2(1 + \gD/d_m)}
      \left\{ 1 + \left[1 + 4\frac{\gD_2}{d_m}\left(1 + \frac{\gD}{d_m}\right)\right]^{1/2} \right\}
\auqe
In the absence of anisotropy energy, namely when only the DDI is taken into account,
the two systems characterized by $(d_m, \gD)$ and $(d_{m2}, \gD_2)$ are
similar and therefore present the same magnetization curve in terms of the reduced field $h$.
Furthermore this holds also whatever the value of $\ep_K$
in the vicinity of zero external field because for random distribution of easy axes
the linear susceptibility $\chi$ does not depend on $\ep_K$ in the superparamagnetic regime.
Doing the transformation (\ref{d2_delta2}), the actual values of $\{\vec{r}_i\}$ are scaled according to
the value of $(d_m + \gD)$. Our hypothesis of a value of
$\s$ for the reduced diameter distribution to be not (or only negligibly) modified holds rigorously in the
quasi monodisperse case ($\s << 1$). Consequently we shall use in the following the scaling transformation
(\ref{d2_delta2}) only in quasi monodisperse situations.

In the present work we focus on both the reduced 
magnetization per unit magnetic volume in the direction of the external applied field,
\equa{magn_def}
M_r = \frac{M(h)}{M_s} = \frac{<\sum_i v(i)\hat{m}_i\hat{h}>}{\sum_i v(i)}
\auqe
and the linear susceptibility,
\equa{chi}
\chi   = \frac{\partial M}{\partial H} 
       = \frac{M_s}{H_{ref}}\left(\frac{d_m}{d_{ref}}\right)^3 \frac{\partial M_r}{\partial h}
       = \frac{M_s}{H_{ref}}\left(\frac{d_m}{d_{ref}}\right)^3 \chi_r      
\blanc \textrm {with} \blanc\chi_r = \frac{\partial M_r(h)}{\partial h}
\auqe
where we have used equation  (\ref{ener_dip2}) to introduce the reduced susceptibility, $\chi_r$. 
The susceptibility can also be obtained from the fluctuations :
\equa{chi_r} 
\chi_r = \beta^*\sum_i \frac{v(i)}{v(d_m)} \left( \frac{< \left(\sum_i v(i)\hat{m}_i\hat{h}\right)^2>}{(\sum_i v(i))^2} - M_r^2 \right). 
\auqe
As a rule, we use this second way with the direct derivative merely used as a check of the calculation.

When the anisotropy energy is zero, the magnetization curve can be simulated either starting from 
$h~=~0$ and increasing the field step by step or from the starurated situation, and decreasing $h$ down to
$h~=~0$. When the anisotropy energy is included and since we may get an opening of the hysteresis loop, we 
start from the saturated case at sufficiently high applied field, and decrease the field beyond $-h_{irr}$ where 
the irreversible field $h_{irr}$ is defined as the value of $h$ below which the hysteresis cycle opens.
In cases where the hysteresis cycle opens, we also define an anhysteretic magnetization curve from the downward 
and the upward magnetization curves which because of the symmetry of our system reads
\equa{anhyst_m}
M_r^{(an)}(h) = \frac{1}{2} (M_r^{(d)}(h) + M_r^{(u)}(h)) = \frac{1}{2} (M_r^{(d)}(h) - M_r^{(d)}(-h))
\auqe

The magnetization curves $M(h)$ in terms of the external field are determined from Monte Carlo 
simulations, by fixing the locations of the particles in the cluster. We consider free boundary 
conditions, and the clusters includes {\it ca} 1000 particles. The dipolar coupling parameter is determined
from equation  (\ref{ener_dip2}). 
In section \ref{comp_exp} we consider a given set of experimental results in order to illustrate 
the model; nevertheless we do not restrict this latter only to this well specified set of samples
but instead use the characteristics of maghemite as typical example for MNP assemblies.
For the bulk magnetization $M_s$ we use a commonly accepted value for
maghemite. Using $M_s$ varying from 80 to 84 emu/g, or $\sim$ 75.0 emu/g,
if we take into account the temperature dependence, and $\rho$ = 4.870g/cm$^3$ leads to
$\mu_0 M_s$ from 0.459 T to 0.514 T; at $T_0$ = 300K
we get $\ep_d^{(0)}$ = 1.0 for $d_{ref}$ varying from 9.665 $nm$ to 10.422 $nm$ and we
use in the following except otherwise mentioned $d_{ref}$ = 10 $nm$ which corresponds to $\mu_0 M_s$
 = 0.488T and $H_{ref}$ = 16.20 kA/m. 
The anisotropy constant $K_1$ cannot be taken equal to the bulk effective magnetocrystalline 
anisotropy constant $K_b$ as it is found to be much larger when the particle size decreases. 
A rather wide spectrum of values can be found in the
literature for $K_1$, corresponding to $(K_1/K_b)$ 
lying in between $\sim$ 4 to 15 for particle diameters of {\it ca} 12 $nm$ or smaller 
\cite{gazeau_1998, fiorani_2002, papaefthymiou_2009, pereira_2010, levy_2011}. In the following we use either 
$(K_1/K_b)$ $\simeq$ 4 or 2, since we consider particles with mean diameters larger than 10 $nm$. 
With $K_b~=~0.47\;10^4~J/m^3$ \cite{fiorani_2002, tronc_2003} this leads to 
$\ep_K^{(0)}$ = 2.38 and 1.19 respectively. 
In any case, both $M_s$ and $K_1$ are to be understood merely as realistic instead of 
truly accurate experimental values given the simplicity of the model. 

Our Monte Carlo simulations are based on the usual Metropolis scheme \cite{binder_1997, allen_1987}; 
the averages are taken over 10 to 40 independent runs each of which consists
in 10$^4$ to 2.10$^4$ MC steps (MCS) of equilibration followed by 2.10$^4$ to 3.10$^4$ 
MCS for the averages calculations. Each MCS consists in one trial move per moment in average.
The trial move on the unit moment $\hat{m}_i$ consists in moving $\hat{m}_i$ to 
$(\hat{m}_i + w\vec{u})/\abs{\hat{m}_i + w\vec{u}}$ where $\vec{u}$ is a random vector picked within the
unit sphere with uniform probability density. This remains to move $\hat{m}_i$ in a cone of maximum
deviation $\delta\theta$ whose value is controlled by the amplitude parameter, $w$.
For $\delta\theta\;<<\;1$, we have $\delta\theta\;\simeq\;w$. The value of $w$ can be either
fixed for a time scale mapping of the MCS or determined in a self consistent way in
order to optimize the sampling by imposing a value for the acceptance ratio, $R$.
The former version of this scheme
corresponds to the time quantified Monte Carlo algorithm (TQMC) \cite{nowak_2000, cheng_2006a}
in its first formulation ignoring the precessional step \cite{nowak_2000}.
In the absence of anisotropy energy, the time scale mapping is irrelevant for the present purpose
since we expect neither a ferromagnetic behavior nor a metastable blocked regime. Thus in this
case, $w$ is self consistently determined in such a way that $R\;=\;0.5$.
Conversely, when $\ep_K^{(0)}\;\neq\;0$, especially for polydisperse distributions we 
expect the largest particles to be in blocked state leading to a remanent state all the more that the
DDI increase the blocking temperature. Hence, especially in the vicinity of $h~=~0$, we deal with 
a metastable state whose life time must be comparable to the long scale measuring time $\tau_m$. 
Strictly speaking one has to perform MC simulations corresponding to $\tau_m$ and to use 
the version of the scheme outlined above allowing a mapping of the MC step on the true relaxing time.
Since we are interested only in the long time behavior (corresponding to the SQUID measurements time
scale), we do not focus on a precise mapping of the MCS scaling time. Instead, we determine
$w$ from the behavior of the instantaneous polarization $M(t)$, versus $t$ in MCS
along a MC run at $h\;=\;0$ starting from $\{\hat{m}_i\}\;=\;\hat{z}$. In other words, 
we chose $w$ in order to avoid nonphysical jumps over the anisotropy energy barrier.
By varying $w$ we get as expected a $w$ dependent evolution of $M(t)$ 
before reaching a fluctuating behavior around a well defined plateau; the long time 
mean value $<M(t)>$ determined beyond some threshold $t$ value and for $t$ 
up to 2.10$^5$ MCS is found independent of $w$ at least for $w$ varying in the range 
$w\;=$ 0.03 to 0.25 for typical values of the parameters we consider ($\ep_d\;\simeq$ 
2 to 8, $\ep_K^{(0)}\;\simeq$ 2.3) and the polydispersity deviation $\s\;=$ 0.28.
Therefore, in the following, we fix $w$ = 0.25 when $\ep_K^{(0)}\;\neq\;0$.

\section {Results and discussion}
\label{result}
\subsection {Weak coupling case}
\label {weak_coupl_tpt}
Before focusing on the powder like situation characterized by a moderate to 
strong dipolar coupling, we consider the weak coupling limit of the DDI,
$\ep_d < 1$ with $\ep_K = 0$ where one can compare the results to the analytical 
one obtained from the thermodynamic perturbation theory and make the link with the 
mean field approximation. The important point is that one can deduce at least 
qualitatively when $\ep_d$ deviates from the limit $\ep_d << 1$, the general 
behavior of the magnetization with respect to the DDI. In this framework, we can 
expand both the magnetization $M_r(h)$ and the susceptibility $\chi_r$ in terms 
of $\ep_d$ \cite{jonsson_2001, margaris_2012}. 
\equa{tpt_1}
M_r(h) &=& M_r^{(0)}(h) + M_r^{(1)}(h) \ep_d + \frac{1}{2}M_r^{(2)} \ep_d^2 \nonumber \\
\chi_r &=& \chi_r^{(0)} + \chi_r^{(1)} \ep_d + \frac{1}{2}\chi_r^{(2)} \ep_d^2
\auqe
$M_r^{(0)}$ and $\chi_r^{(0)}$ correspond to the non interacting case, namely
\equa{lang_1}
 M_r^{(0)}(h) &=& \frac{\int v(d)L(\beta^*(d/d_{m})^3 h) f(d) d(d)}{\int v(d) f(d) d(d)}  \nonumber \\
\auqe
where $L$ is the Langevin function; $\chi_r^{(0)}$ is directly related to $M_r^{(0)}(h)$ and 
at $h~=~0$ leads to the linear susceptibility of the non interacting system
\equa{chi_no_ddi}
\chi_r^{(0)}(0) = \beta^*(d_6^*/d_3^*)/3 ~=~\beta^*\exp(27\s^2/2)/3
\auqe
where $d_s^*$ is the $s$-th reduced moment of the distribution $f(d)$.
Equation (\ref{chi_no_ddi}) explicits the effect of the polydispersity through the factor $(d_6^*/d_3^*)$, 
written here in terms of $\s$ for the lognormal distribution.
The expansions (\ref{tpt_1}) which have been explicited in the framework of the TPT in
\cite{jonsson_2001, kachkachi_2005} depend on geometrical sums which can be directly calculated
from the structure considered. Moreover, the linear part with respect to $h$ of $M_r^{(1)}(h)$ can be
deduced in the mean field approximation of the magnetization
which introduces the DDI contribution to $M_r$ from the demagnetizing field and  follows from
$M_r(H_a)~=~M_r^{(0)}(H_{eff})$ where $M_r^{(0)}$ corresponds to the non interacting system and
$H_{eff}$ is the effective field
\equa{mf_1}
  H_{eff}   = H_a -(D_\alpha - 1/3)M_v^{(0)}(H_a) ~;~ 
\auqe
$D_\alpha$ is the demagnetizing factor of the sample in the direction of the external
field, $\hat{h} = e_{\alpha}$, and $M_v$ is the total magnetization per unit volume which 
is related to either the number of MNP per unit volume, $\rho$, or the MNP volumic fraction, $\varphi_v$,
through
\equa{mf_1b}
M_v^{(0)} = M_s M_r^{(0)}(H_a) \rho \int v(d)f(d) = M_s M_r^{(0)}(H_a) \varphi_v
\auqe
Using equation (\ref{lang_1}) for $M_r^{(0)}$ and keeping only the first order term with respect to $h$, we get
\equa{mf_1c}
H_{eff} = H_a - \frac{1}{3}(D_\alpha - 1/3)M_s\beta^*\frac{d_6^*}{d_3^*}h
\auqe
which is then inserted in the mean field expression for $M_r(H_a)$;
then form an expansion of $M_r^{(0)}$ at first order with respect to $h$ 
and from equation (\ref{ener_dip2}) for $\ep_d$, we get
\equa{mf_2}
M_r^{(1)}(h) = -\beta^{*2}(D_\alpha - 1/3)\frac{8}{3}\varphi_v \left(\frac{d_6^*}{d_3^*}\right)^2 h
\auqe
Equation (\ref{mf_2}) can
be equivalently rewritten, in terms of $\Delta M_r = M_r(\ep_d) - M_r(\ep_d=0)$,  as
\equa{mf_3}
   \frac{\partial (\Delta M_r(h))}{\partial \ep_d} &=& 
             -\beta^{*2}(D_\alpha - 1/3)\frac{8}{3}\varphi_v \left(\frac{d_6^*}{d_3^*}\right)^2 h
   \nonumber \\   {\textrm and}~   C_2                  &= &
             \frac{\partial^2 \Delta M_r(h)}{\partial \ep_d \partial h}(h = 0, \ep_d = 0) 
             \equiv \frac{\partial \chi_r}{\partial \ep_d} ~=~ 
            -\beta^{*2} (D_\alpha - 1/3)\frac{8}{3}\varphi_v \left(\frac{d_6^*}{d_3^*}\right)^2
    \nonumber \\
\auqe
A result in agreement with refs. \cite{jonsson_2001} and \cite{kachkachi_2005} in the
monodisperse case. Here, the important point is that we explicitly write down the effect of
the polydispersity through the factor $(d_6^*/d_3^*)^2$ which strongly deviates from unity
once $\s$ takes non negligible values. It is worth mentioning that the preceding equations
hold when either $\ep_d << 1$ or $\varphi_v <<1$. We have performed MC simulations of the magnetization 
at small values of the coupling constant for prismatic clusters corresponding to either
well ordered (simple cubic, and c.f.c) or random structures with a monodisperse particles
distribution, and a random structure with a polydispersity characterized by $\s = 0.28$.
The results for the second derivative of $\Delta M_r$ with respect to $\ep_d$ and
$h$, $C_2$, is displayed in table \ref{tab_pr_lin}. As can be seen, especially for $\hat{h}~=~\hat{z}$,
the mean field approximation or equivalently the linear contribution of the TPT compares well 
with the MC simulations and in particular for the polydispersity effect.

Since spherical or cubic systems are characterized by $D_{\alpha}$ = 1/3, this first term vanishes
in these situations and one is left for the DDI contribution with $\Delta M_r \propto \ep_d^2$
and similarly for $\chi_r$. 
Moreover, still for isotropic systems, we know that the DDI contribution to both
$M_r$ and $\chi$ is negative. Therefore the magnetization is all the more reduced due to the DDI
that the coupling constant $\ep_d$ increases.
From the analytical results of the TPT we can calculate the proportionality coefficient relating
$(\chi_r(\ep_d) - \chi_r(0))$ to $\ep_d^2$. We have thus compared the MC simulation to the theoretical
small $\ep_d$ expansion in the simple cubic structure case and a monodisperse distribution. 
From this comparison, see figure \ref{pert_cs_sph}, we can check that the TPT gives an accurate result 
only for $\ep_d < 0.2$ as expected. Furthermore, from a description based on the
dipolar fields distributions which can be seen as a generalization of the mean field type
of approach, ref. \cite{al-saei_2011} have shown also that the dipolar interactions in isotropic systems 
decrease the magnetization. This decreases is related to the non linearity
with respect to the applied field of the non interacting contribution $\chi_r^{(0)}$
to the susceptibility. Notice that this second type of approach, which remains qualitative in the
absence of a theory to deduce the dipolar field distribution, is not restricted to the weak 
coupling case. Hence, as a general rule, we expect that in an isotropic sample
the DDI tend to reduce the magnetization.  
However, this reasoning does not hold at high fields where the Zeeman term dominates on the DDI 
and where we expect an approach to saturation, 
close to what is obtained in the non interacting case 
deduced from the high field expansion of $M_r^{(0)}$ of equation (\ref{lang_1}), namely
($M_r(h) \sim 1 - 1/(\beta^*d_3^*h)$). 
\subsection {Spherical clusters in the strong coupling case}
\label {strong_coupling_sph}
We now consider, exclusively for spherical clusters, the moderate to strong dipolar coupling 
case corresponding to the experimental situation of typical coated maghemite NP powders \cite{de-montferrand_2012}, 
with $\ep_d^{(0)} = 1$ for $d_{ref} = 10 nm$ and a coating layer of $\gD = 2.0 nm$. 
The median diameter varies from $d_{ref}$ to $2\times d_{ref}$ and the standard 
deviation of the distribution $ln(d)$ is taken from $\s = 0.05$ to represent the quasi 
monodisperse case to $\s = 0.50$ 
to represent a large polydispersity. The importance of $\s$ on the MNP distribution in the
clusters is clearly seen on figure (\ref{centre_cluster}). 
Notice that a standard value obtained 
experimentally is {\it ca.} 0.20 $\sim$ 0.30 which is represented here by $\s$ = 0.28.
In the first step we neglect the anisotropy contribution ($\ep_K = 0$) and focus only on the DDI.
First of all we analyse the linear susceptibility, $\chi$ which provides the behavior at low 
field of the magnetization. 
Since in our model, with a constant coating layer thickness, $\gD$, the dipolar coupling
constant scales as $(d_m/d_{ref})^3$ we expect in the vicinity of $h \sim 0$ a reduction of the 
magnetization higher for large median diameters, where the initial non interacting magnetization 
$M_r^{(0)}$ is higher. In the quasi monodisperse case, $\s = 0.05$ we make use of the scaling 
transformation introduced in equation (\ref{sum_ddi_2}) to explicit the effect of the coating 
layer thickness $\gD$ on $\chi$ by using only one set of simulations for $\gD/d_{ref} = 0.20$. 
We checked for $\gD\;=\;0.8$ and 2 values of $d_m$ the reliability of this scaling transformation
(see figure (\ref{chi_diam})). Therefore, in the quasi monodisperse  
case we have a rather complete picture of both the effect of the variation of the median diameter, 
$d_m$ and of the distance of closest approach between NP, controlled by the coating layer thickness, 
$\gD$. The result for $\chi$ determined with $\s$ = 0.05 is displayed and compared to the non 
interacting case $\chi^{(0)}$ in figure (\ref{chi_diam}). As expected, when increasing the particle 
size and consequently the DDI coupling constant $\ep_d$, an increasing reduction of $\chi$ is 
obtained. This reduction is of course all the more important that $\gD$ is small. The important 
result is that we can get a plateau, which means that $\chi$ may becomes particle size independent
beyond a threshold value which is, as expected, strongly $\gD$ dependent.
As already mentioned, for random distribution of easy axes, $\chi$ does not depend on 
$\ep_K$ in the superparamagnetic regime, and accordingly this result holds also in the case 
where the anisotropy is included.

The dependence of $\chi$ on both $d_m$ and $\gD$ can be used to deduce the behavior of
$\chi$ with the NP volumic fraction (or concentration) at fixed value of $d_m$ 
through the relation $\varphi_v\;=\;\varphi_m(d_m/(d_m+\gD))^3$ 
with $\varphi_m\;=\;\varphi_v(\gD=0)$. Doing this, in agreement with other MC results
\cite{chantrell_2000, serantes_2010}, we get a monotonous decrease of $\chi$ with the 
increase in $\varphi$, as shown on figure (\ref{phi_c_d}). Furthermore this shows that a fit 
of the NP size on the magnetization curve by using a Langevin function does not hold 
beyond a critical value $\varphi_c$ of the volumic fraction. We can estimate this latter 
by imposing that $\chi/\chi(\ep_d\;=\;0)$ is larger than some threshold value
say $\lambda$, leading the determination of $\varphi_c$ through $\chi(\varphi_c)/\chi(\ep_d\;=\;0) = \lambda$.
The result obtained by using $\lambda$ = 0.80 is displayed on figure (\ref{phi_c_d}). 

The magnetization curves in terms of the reduced external field $H_a/H_{ref}$
for three values of the median diameter, still for $\s$ = 0.05 is shown in figure 
(\ref{m_diam_0.05}) and compared to the non interacting diameter distribution weighted 
Langevin curves. We clearly see the important reduction of the magnetization compared 
to the non interacting case, and the 
very weak dependence 
of the low field behavior with
respect to the median diameter which is expected as the considered sizes 
are either close to the onset of the $\chi(d_m)$ curve plateau corresponding to
$(\gD/d_{ref})$ = 0.20 ($d_m/d_{ref}$ = 1) or pertain to this later ($d_m/d_{ref}$ = 1.33 
and 2.00). On the other hand at low external fields the nearly size independence of the 
magnetization is correlated with a quasi linear behavior of $M_r(h)$ with respect 
to $h$, which seems coherent with the interaction fields distribution description 
\cite{al-saei_2011}.

Then we introduce the polydispersity at fixed values of $d_m$. First we consider the
case $d_m/d_{ref}$ = 1.33, as an example of median diameter located in the plateau region 
of the $\chi(d_m)$ curve. In this case we expect a very weak dependence of the magnetization
with respect to the polydispersity in the low field region and this is confirmed by the MC 
simulations.  
Indeed, we get only small changes of $M_r(h)$ with $\s$ as can be seen in figure 
(\ref{m_sigma_1.33}). The magnetization curves corresponding to $\s$ up to 0.40 are very 
close to each other for the values of the field for which $M_r < 0.70$; beyond this value, 
the deviations between the different magnetization curves reflect mainly the approach to 
saturation where $M_r(h) \sim (1 - 1/(\beta^*d_3^*h))$ depends on $\s$ through $d_3^*$. 
The deviation from the quasi monodisperse situation over the whole field range occurs 
for $\s \geq$ 0.5.
Conversely, when the median diameter is taken outside of the $\chi(d_m)$ plateau, 
as is the case for $d_m/d_{ref}$ = 1.0 the polydispersity has a noticeable influence 
on the magnetization as shown in figure (\ref{m_sigma_1.00}) for $\s$ ranging from
0.05 to 0.40.

In the superparamagnetic regime the MNP anisotropy energy  modifies the magnetization
curve for intermediate values of the field and leads to a reduction of $M_r$ since the
moments tend to be pinned in the easy axes directions. Taking into account $\ep_K$ thus 
reduces further $M_r$ for $h$ between the low field region controlled by the DDI and the 
approach to saturation controlled by the Zeeman energy. 
In the quasi monodisperse case, the blocking temperature corresponds to that of the median 
diameter, namely for non interacting particles, $k_BT_b \simeq K_1v(d_m)/25$ 
or equivalently for the reduced blocking temperature 
$1/\beta_b^* \simeq \ep_K^{(0)}(d_m/d_{ref})^3/25$ leading to $1/\beta_b^* \simeq$ 0.225 
for $d_m/d_{ref}$ = 1.33. Here we restrict to the room temperature, $\beta^*$ = 1, 
and we expect the system to be in the superparamagnetic regime even for short times. 
Indeed for $\s$ = 0.05 our MC simulations confirm the superparamagnetic regime. 
The result is displayed and compared to the $\ep_K$ = 0
case in figure (\ref{anis_0.05_1.0-1.33}) for $d_m/d_{ref}$ = 1.0 and 1.33. As expected, 
the anisotropy energy does not affect the $M(H)$ curve in the vicinity of $H=0$ due to the 
random distribution of easy axes. Moreover, when $d_m/d_{ref}$ = 1, the $M(H)$ curve for 
intermediate values of the field is only weakly modified by the anisotropy energy while for 
$d_m/d_{ref}$ = 1.33 a noticeable deviation is obtained.

The influence of the polydispersity on the magnetization curve when the anisotropy energy
is included is shown for $d_m/d_{ref}$ = 1.33 on figure (\ref{anis_1.33_0.05-0.35}) for 
$\s$ ranging from 0.05 to 0.35. Because of the largest particles in the distribution,
the system is no more in the superparamagnetic regime for the MC runs considered up to
10$^5$ MC steps. On the qualitative point of view this is expected since 
$1/\beta_b^*$ behaves as $d^{*3}$ in the absence of DDI and moreover increases with the DDI.
As a result an opening of the hysteresis cycle is obtained with remanence magnetization
and coercive field increasing with $\s$ as shown in figure (\ref{anis_1.33_0.05-0.35})
in the particular case $d_m/d_{ref}$ = 1.33.
The magnitude of the hysteresis cycle opening is expected to increase with $d_m/d_{ref}$
and is indeed found very weak for $d_m/d_{ref}$ = 1.0.
The determination of the remanence in terms of $d_m$ and the measuring time
is beyond the scope of this work; we nevertheless note (see section \ref{comp_exp})
that the hysteresis cycle opening for large values of $d_m$ is in qualitative 
agreement with experiment.
\subsection {Comparison with experiment}
\label{comp_exp}
We now consider experimental results obtained recently on $\gamma-Fe_2O_3$ NP powders
samples differing by their size \cite{de-montferrand_2012}. 
The experimental protocol for the synthesis is described in \cite{de-montferrand_2012}. 
The particles are coated with (5-hydroxy-5,5-bis(phosphono)pentanoic acid)
which provides a coating layer of thickness ${\it ca}$ 
2 nm between particles. As a result of the synthesis method, the standard deviation of the
diameter distributions as determined by TEM takes nearly the same value in the 4 samples
considered, namely, $\s~\simeq~0.26$. The saturation magnetization $M_s$ is found to be in
between 61 and 70 $emu/g$ for the distributions characterized by $d_m$ = 10, 12, 18 and 21 $nm$.
Although these values are smaller than the bulk value at room temperature ($\sim$ 75 $emu/g$)
the difference is small enough for the spin canting to be neglected in first approximation.
It is worth mentioning that the magnetic properties of these NP assemblies
have been measured also in diluted solution and, although the possible formation of 
clusters and/or chains in the presence of the external field cannot be ruled out, this
allows for an estimation of the interaction effect. 
When going from the dispersed samples to the powder ones, we observe both a strong 
reduction of the magnetization and its weak size dependence in the low field region 
\cite{de-montferrand_2012}. 
According to our simulations, both effects result from the DDI. 
The DDI induced reduction of $M_r(H)$ in the absence of demagnetizing effects 
is a general simulation result 
\cite{chantrell_2000, garcia-otero_2000, serantes_2010} and a similar
trend has been obtained experimentally \cite{gonzales-weimuller_2009},
and can be deduced from the FC/ZFC measurements in the superparamagnetic regime of
either bare or Si coated $\gamma-$Fe$_2$O$_3$ NP \cite{pereira_2010}.
Beside its rather weak size dependence the other feature of
the experimental reduced magnetization curves, $M_r(H)$ in the
low field region (see  figure (\ref{m-h_exp_lf}) is the opening of the hysteresis
cycle for the largest sizes beyond $d_m$ = 12 $nm$. These two points are in 
qualitative agreement with the MC simulations on our model although the opening of the cycle 
becomes noticeable for larger median diameters ($d_m~\sim~18~nm$) than in MC simulations.

In the present work, we do not compare the experimental magnetization curve in the whole
range of field with the results of either a mean field approach or the TPT. In any case
the values of the dipolar coupling corresponding to the experimental powders samples
($\ep_d^{(eff)}~\sim~$ 0.6 to 6.0 when $\ep_d \sim$ 1 to 8 and the effect of $\gD$ 
is taken into account) fall outside of the range of validity of the TPT. Indeed this
later is limited to {\it ca} $\ep_d <$ 1/6 according to ref. \cite{jonsson_2001},
the analytical approach based on TPT of ref. \cite{margaris_2012} is shown to be very 
accurate for $\ep_d~<$ 0.25 and valid for $\ep_d~<$ 0.50 in the monodisperse case
and in section \ref{weak_coupl_tpt} we found that $\chi$ as calculated from the TPT start
to deviate from the simulated results at $\ep_d \sim$ 0.2.
Moreover the accounting of the polydispersity is expected to worsen the lack of
accuracy of the TPT with the increase of $\ep_d$.

In figure (\ref{comp_exp_d10}) we compare the experimental and simulated $M_r(H)$ 
for the applied field in the low to intermediate range for $d_m\;=\;10\;nm$. 
The agreement is quite satisfactory up to $H$ = 60 kA/m where $M_r\;\simeq\;0.7$. 
Then for median diameters $d_m~>~$ 12 nm ($d_m/d_{ref}~>~1.2$, we get for both the experimental samples and the MC
simulations an opening of the hysteresis cycle.
However, as can be deduced from figures (\ref{m-h_exp_lf}) and (\ref{anis_1.33_0.05-0.35})
the irreversible field, $H_{irr}$ is found much larger in the MC simulations than in the experiment.
Notice that we do not try to map rigorously the MC time scale to the actual measurement time $\tau_m$,
and this is plays a central role on this point. Therefore concerning the MC simulations we consider
in the following
the anhysteretic magnetization as defined in equation (\ref{anhyst_m});
using this later remains to ignore the hysteresis cycle ({\it i.e.}
the remanence and the coercive field) or to consider the infinite time scale limit.
We compare the experimental $M(H)$ to the simulated ones in figures (\ref{comp_exp_d12},
\ref{comp_exp_d20}) for the median sizes $d_m$ 12 and 20 nm respectively. 
Given the simplicity of the model which does not include at all any structure in the particles, 
and the absence of fitting parameter 
the agreement is qualitatively satisfactory when $\ep_K^{(0)}\;=\;2.38$
in particular for the overall variation of $M_r(H)$ at low fields ($H\;<\;40$ kA/m).

Since in this range of fields, the deviation of $M_r$ from the non interacting case
is governed by the DDI, we can conclude that the strong reduction of the $M_r(H)$ variation
with respect to $H$ and its relative size independence when compared to the diluted 
solutions counterpart is the DDI signature.
For median sizes larger than $d_m\;=\;10\;nm$, the main discrepancy between the 
simulated and experimental $M_r(H)$ curves is the strong non linearity in the very
vicinity of $H\;=\;0$. This is clearly due to the oversimplification of the model
in which the particles are uniform single domain ones.

\subsection {Conclusion}
\label {conclusion}
In this work we have used MC simulations to investigate both the median size and
polydispersity effects on the magnetization curve of densely packed clusters of 
single domain magnetic NP. 
An important result is the plateau in the $\chi(d_m)$ curve
in the quasi monodisperse case for small values of the coating layer $\gD$, 
which emphasizes the much reduced size dependence of the $M(H)$ low field
dependence in the concentrated systems.
Despite of the simplicity of the model, some important 
features of the experimental $M(H)$ on powder samples are reproduced, especially
concerning the DDI signature which occurs principally at low fields and its dependence on the particle size.
In order to get a satisfactory agreement with experiments, it appears that
the internal polarization structure of the NP should be introduced.

\section* {Acknowledgements}
\label {ackno}
This work was granted access to the HPC resources of CINES under the allocation 
2012-c096180  made by GENCI (grand Equipement National de Calcul Intensif).

\section*{Tables}
\begin{figure}[H] 
\includegraphics[width = 0.75\textwidth, angle = -00.00]{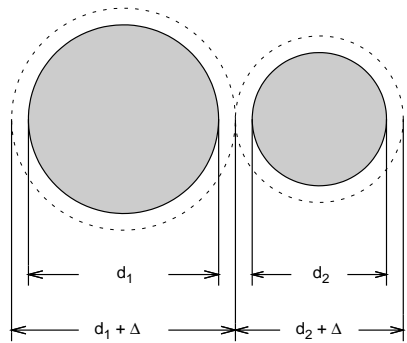}
\caption{\label{shema_contact}
Shematic view of the configuration for two particles coated by the layer of thickness
$\gD/2$ at contact.
}
\end{figure}

\begin{figure}[H] 
\includegraphics[width = 0.210\textwidth, angle = -00.00]{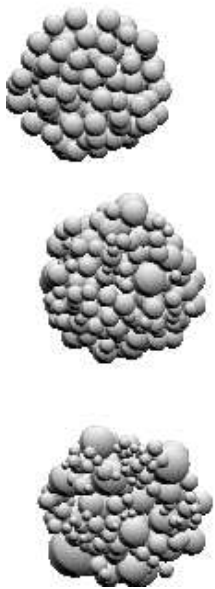}
\caption{\label{centre_cluster}
Central part of the clusters corresponding to $d_m/d_{ref}$ = 1.33 and
$\s$ = 0.05, 0.28 and 0.50 from top to bottom.
}
\end{figure}

\begin{figure}[H] 
\includegraphics[width = 0.75\textwidth, angle = -00.00]{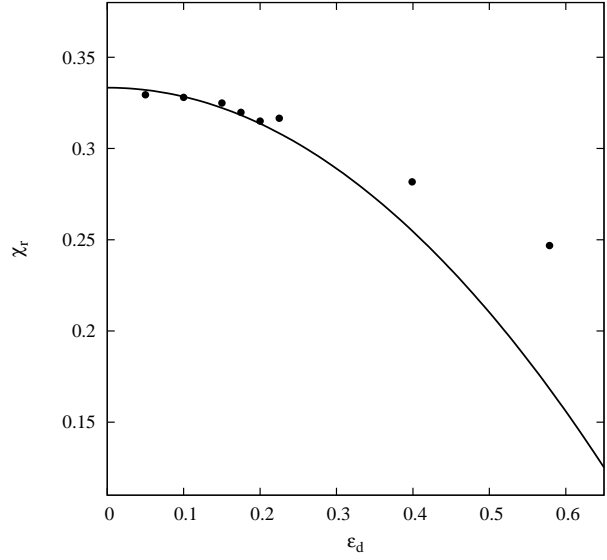}
\caption{\label{pert_cs_sph}
Comparison of $\chi_r$ in terms of $\ep_d$ as calculated from the TPT of Ref. \cite{garcia-palacios_2000},
(solid line) and the present MC simulation (symbols) for a spherical cluster of simple cubic
structure with $N_p$ = 1021 particles and a monodisperse distribution. $\beta^*$ = 1.0.
}
\end{figure}

\begin{figure}[h,t] 
\includegraphics[width = 0.75\textwidth, angle = -00.00]{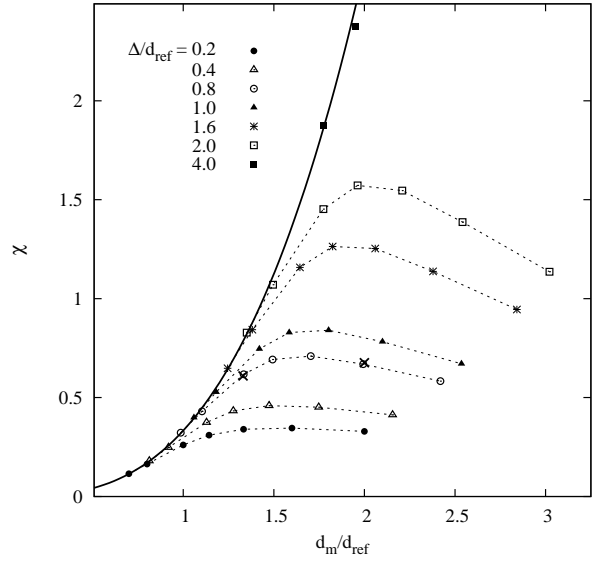}
\caption{ \label{chi_diam} 
Linear susceptibility in terms of the median size $d_m/d_{ref}$
for different values of the coating layer thickness $\gD$ in the quasi monodisperse
case, $\s$ = 0.05 and $\beta^*$ = 1.0.
The two crosses on the $\gD/d_{ref}$ = 0.8 curve correspond to the direct calculation
without using the scaling transformation (\ref{d2_delta2}). The dotted lines are guides to the eye 
and the solid line corresponds to the non interacting case.
} 
\end{figure}

\begin{figure}[h,t] 
\includegraphics[width = 0.75\textwidth, angle = -00.00]{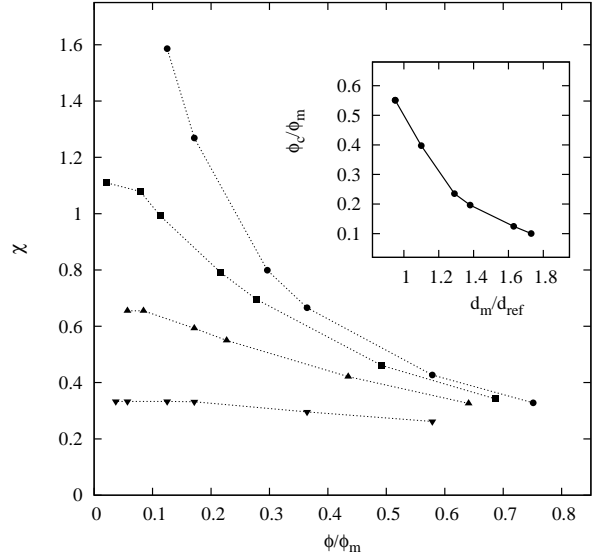}
\caption{ \label{phi_c_d} 
Susceptibility $\chi$ in terms of the reduced volumic fraction, for 
$\s$ = 0.05, $\beta^*$ = 1.0, and  
$d_m/d_{ref}$ = 2.0 (solid circles); 1.50 (solid squares); 1.25 (upward triangles) and 1.0 (downward triangles).
$\varphi_m$ = $\varphi(\gD = 0)$. In the present work $\varphi_m\;\simeq$  0.59.
Inset: reduced critical volumic fraction defined as $\chi(\varphi,\ep_d)/\chi(\ep_d=0)$ = 0.80
in terms of the median particle size. 
} 
\end{figure}

\begin{figure}[h,t] 
\includegraphics[width = 0.75\textwidth, angle = -00.00]{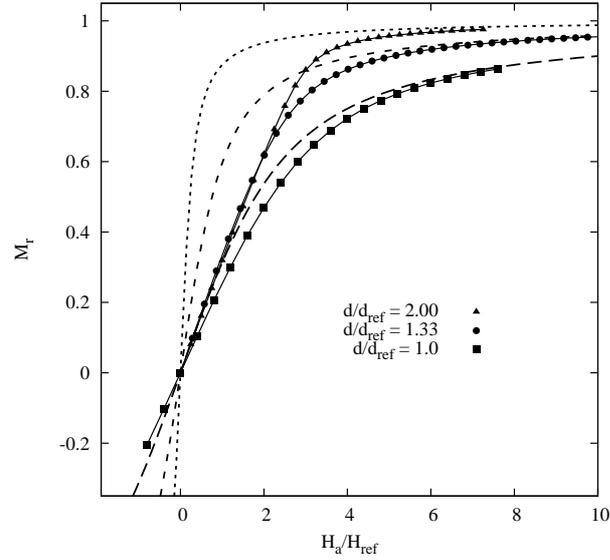}
\caption{ \label{m_diam_0.05} 
Magnetization in terms of the applied field for different values of the median diameter,
$\gD/d_{ref}$ = 0.20, $\beta^*$ = 1.0 and $N_p$ = 1007 in the quasi monodisperse case, $\s$ = 0.05.
The corresponding non interacting curves (diameter distribution weighted Langevin curves)
for $d_m/d_{ref}$ = 1.0 (long dash), 1.33 (short dash) and 2.0 (dotted line)
are displayed for comparison.
} 
\end{figure}

\begin{figure}[h,t] 
\includegraphics[width = 0.75\textwidth, angle = -00.00]{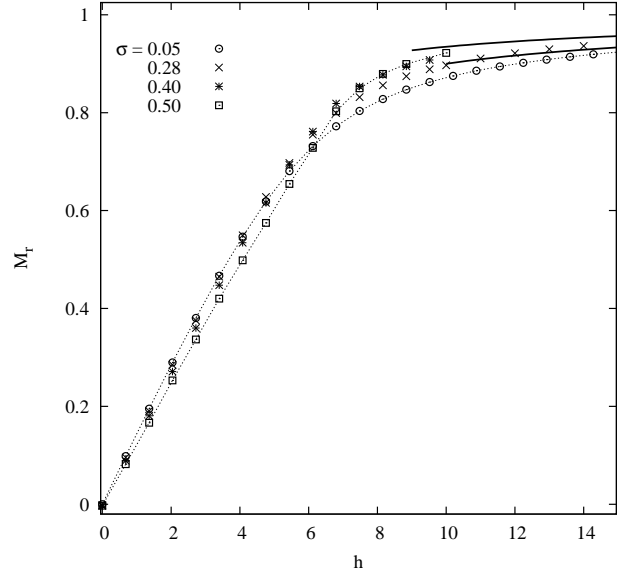}
\caption{ \label{m_sigma_1.33} 
Magnetization in terms of the applied field for different values of the $ln(d)$ standard
deviation $\s$ and $d_m/d_{ref} = 1.33$, $\gD/d_{ref}$ = 0.20 and $\beta^*$ = 1.0.
$N_p$ = 1007 ($\s$ = 0.05), 923 ($\s$ = 0.28); 985 ($\s$ = 0.40) and 990 ($\s$ = 0.50).
The dotted lines are quides to the eye.
The solid lines are the asymptotic limits for $\s$ = 0.05 (bottom) and $\s$ = 0.50 (top).
} 
\end{figure}

\begin{figure}[h,t] 
\includegraphics[width = 0.75\textwidth, angle = -00.00]{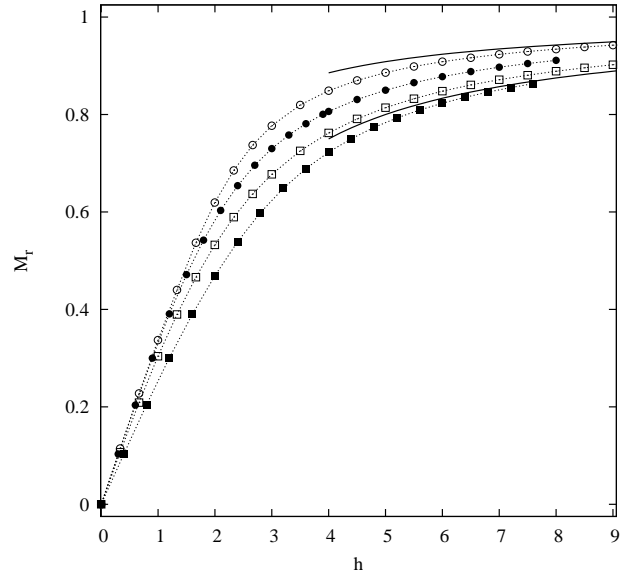}
\caption{ \label{m_sigma_1.00} 
Magnetization in terms of the applied field for 
$\s$ = 0.05, 0.20, 0.28 and 0.40 from bottom to top.
and $d_m/d_{ref} = 1.00$, $\gD/d_{ref}$ = 0.20, $\beta^*$ = 1.0.
The dotted lines are guides to the eyes and the solid lines are the corresponding asymptotic 
limits for $\s = 0.05$ and $\s = 0.40$.
} 
\end{figure}

\begin{figure}[h,t] 
\includegraphics[width = 0.75\textwidth, angle = -00.00]{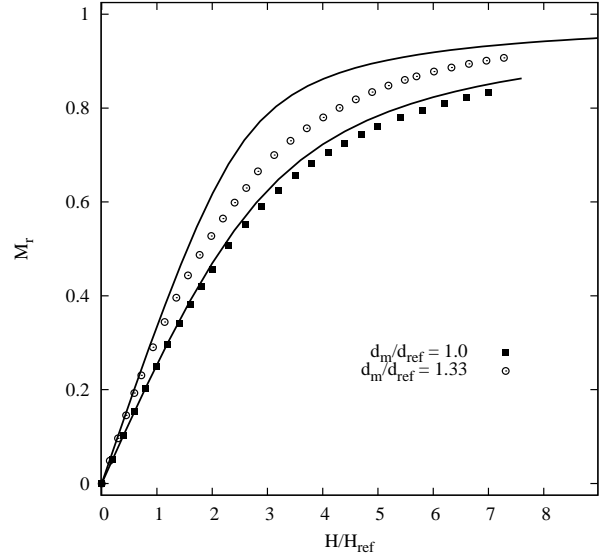}
\caption{ \label{anis_0.05_1.0-1.33} 
Magnetization curve with the anisotropy
energy $\ep_K^{(0)}$ = 2.38, $\s$ = 0.05 and the particles sizes as indicated. 
The solid lines correspond to $\ep_K$ = 0. $\beta^*$ = 1 and $N_p$ = 1007.
} 
\end{figure}

\begin{figure}[h,t] 
\includegraphics[width = 0.75\textwidth, angle = -00.00]{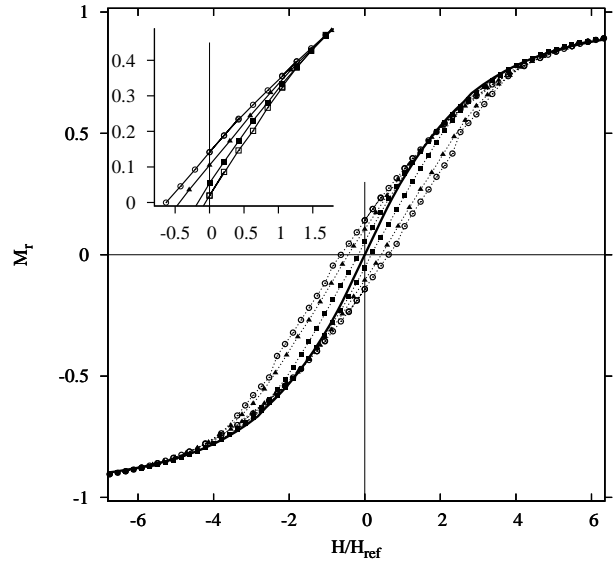}
\caption{ \label{anis_1.33_0.05-0.35} 
Magnetization curve with the anisotropy
energy $\ep_K^{(0)}$ = 2.38, $d_m/d_{ref}$ = 1.33 and $\beta^*$ = 1 in the polydisperse case 
with $\s$ = 0.35 (open circles); 0.28 (triangles); 0.20 (squares)
compared to the quasi monodisperse case, $\s$ = 0.05 (solid line).
Insert: detail of the downward magnetization curve  in the vicinity of $h$ = 0, showing 
the evolution of the remanence and
coercivity with $\s$ for $\s$ = 0.35, 0.28, 0.20 and 0.10 (open squares)  from top to bottom.
} 
\end{figure}

\begin{figure}[h,t] 
\includegraphics[width = 0.75\textwidth, angle = -00.00]{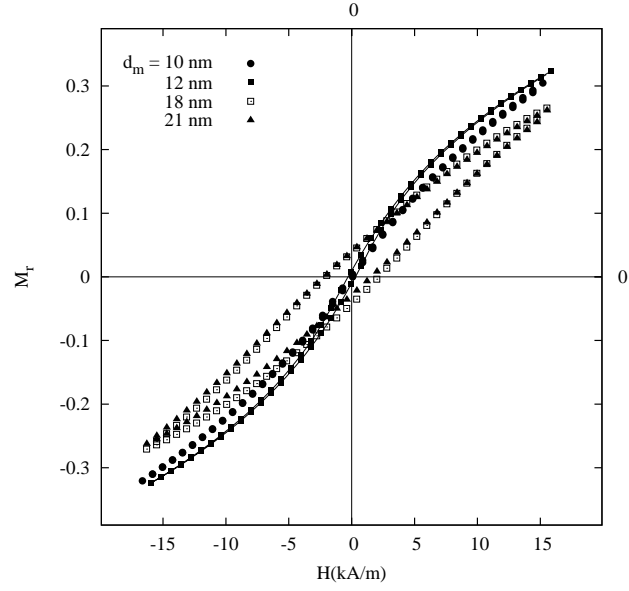}
\caption{ \label{m-h_exp_lf} 
Experimental low field magnetization curves at room temperature for powder samples of 
$\gamma-$Fe$_2$O$_3$ from ref. \cite{de-montferrand_2012} with polydispersity 
$\s\;\simeq$ 0.26 and median sizes as indicated.
} 
\end{figure}

\begin{figure}[h,t] 
\includegraphics[width = 0.75\textwidth, angle = -00.00]{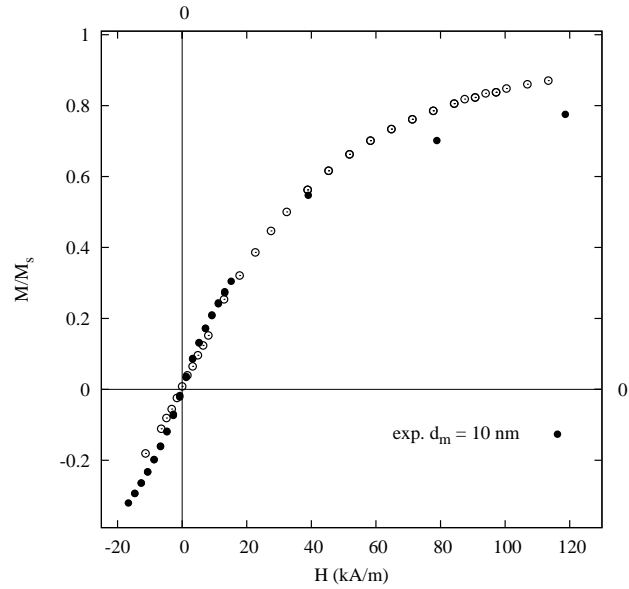}
\caption{ \label{comp_exp_d10} 
Comparison between MC simulations and the experimental magnetization curve at 
room temperature for $d_m$ = 10 $nm$. The simulation are performed with $\beta^*$ = 1, 
$\ep_K^{(0)}$ = 2.38 and $N_p$ = 923.
} 
\end{figure}

\begin{figure}[h,t] 
\includegraphics[width = 0.75\textwidth, angle = -00.00]{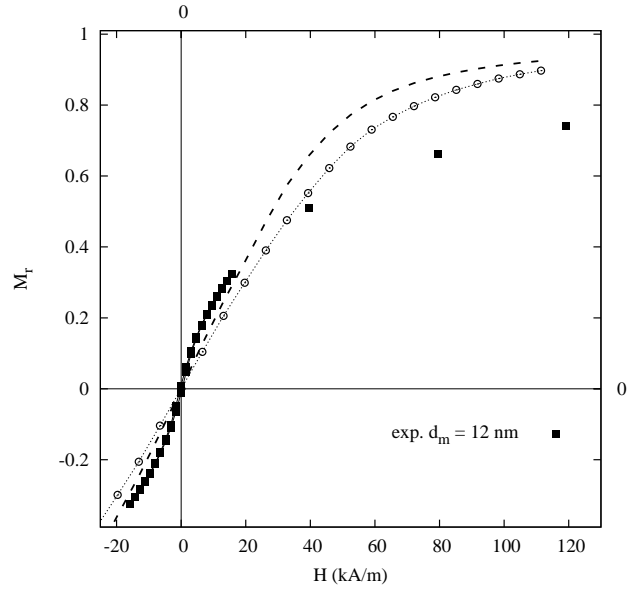}
\caption{ \label{comp_exp_d12} 
Comparison between MC simulations and the experimental magnetization curve at 
room temperature for $d_m$ = 12 $nm$. The simulations are performed with $\beta^*$ = 1, 
$N_p$ = 980, $\ep_K^{(0)}$ = 2.38 (open circles) or 1.19 (dashed line).
The dooted line is a guide to the eye.
} 
\end{figure}

\begin{figure}[h,t] 
\includegraphics[width = 0.75\textwidth, angle = -00.00]{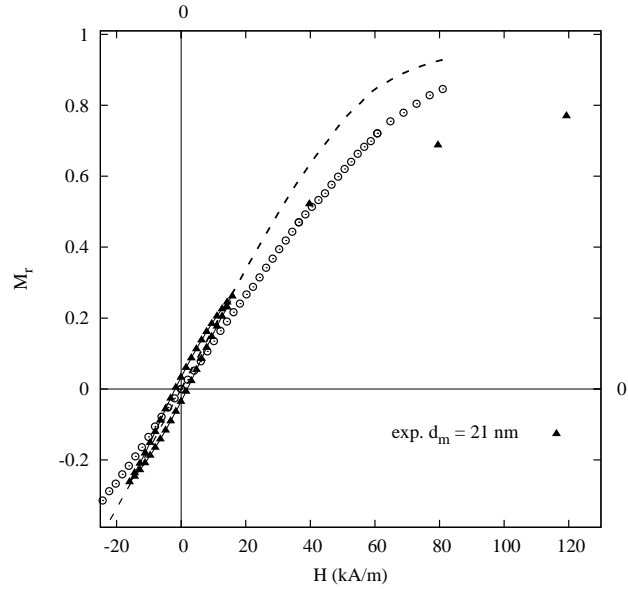}
\caption{ \label{comp_exp_d20} 
Comparison between MC simulations and the experimental magnetization curve at 
room temperature for $d_m$ = 21 $nm$. The simulations are performed with $d_m$ = 20 nm,
$\beta^*$ = 1, $N_p$ = 998, $\ep_K^{(0)}$ = 2.38 (open circles) or 1.19 (dashed line).
} 
\end{figure}

\begin{table}[h]
\begin{tabular}{ | c || c | c | c | c | }
\hline
  struct. 	   & cs$^{(a)}$  &  cfc$^{(b)}$	  &  ~rand. $\s$ = 0$^{(c)}$ ~& ~rand. $\s$ = 0.28$^{(d)}$~ \\
\hline
  $\varphi_v$       & $\pi/6$     &   0.74   &  0.525   &  0.580  \\
\hline
 ~$C_2^{(f)}$~ 	   & ~0.2775~    & ~ 0.3367~& ~0.2534~ &  ~1.7934~ \\ 
 ~$C_2^{(g)}$~ 	   & -0.4745     &  -0.6109 & -0.4390  & -3.1506 \\ 
\hline
 ~$C_2^{(h)}$~     &  0.2217 &  0.3133 &  0.2217        &  1.6173 \\
 ~$C_2^{(i)}$~     & -0.4433 & -0.6265 & -0.4433        & -3.2354 \\
\hline
\end{tabular}
\caption {\label{tab_pr_lin}
Second derivative  $C_2$ = $\partial^2(\Delta M_r)/\partial\ep_d \partial h$ at $\ep_d=0$ and $h = 0$
for a primatic cluster of 1024$^{(a)}$, 1103$^{(b)}$, 1054$^{(c)}$ and 879$^{(d)}$
particles, characterized by $L_x = L_y$ and $L_z = L_x/5$. 
MC simulation with $\hat{h} = \hat{x}~^{(f)}$, $\hat{h} = \hat{z}~^{(g)}$;
equation (\ref{mf_3}) with $\hat{h} = \hat{x}~^{(h)}$, $\hat{h} = \hat{z}~^{(i)}$.
$~^{(h),\;(i)}$ The demagnetizing factor entering 
equation (\ref{mf_3}) is taken from \cite{aharoni_1998} and the moments $d^*_s$ are taken from the
actual diameters distribution of the cluster considered. 
}
\end{table}
\newpage
%
%
%
%
\newpage
%
\begin{singlespace} 
\vskip 0.0 cm \noindent 
{\bf Figure captions} 
\begin{itemize} 
\item[Figure \ref{shema_contact}]
Shematic view of the configuration for two particles coated by the layer of thickness
$\gD/2$ at contact.
\item[Figure \ref{centre_cluster}]
Central part of the clusters corresponding to $d_m/d_{ref}$ = 1.33 and
$\s$ = 0.05, 0.28 and 0.50 from top to bottom.
\item[Figure \ref{pert_cs_sph}]
Comparison of $\chi_r$ in terms of $\ep_d$ as calculated from the TPT of Ref. \cite{garcia-palacios_2000},
(solid line) and the present MC simulation (symbols) for a spherical cluster of simple cubic
structure with $N_p$ = 1021 particles and a monodisperse distribution. $\beta^*$ = 1.0.
\item[Figure \ref{chi_diam}] 
Linear susceptibility in terms of the median size $d_m/d_{ref}$
for different values of the coating layer thickness $\gD$ in the quasi monodisperse
case, $\s$ = 0.05 and $\beta^*$ = 1.0.
The two crosses on the $\gD/d_{ref}$ = 0.8 curve correspond to the direct calculation
without using the scaling transformation (\ref{d2_delta2}). The dotted lines are guides to the eye 
and the solid line corresponds to the non interacting case.
\item[Figure \ref{phi_c_d}] 
Susceptibility $\chi$ in terms of the reduced volumic fraction, for 
$\s$ = 0.05, $\beta^*$ = 1.0, and  
$d_m/d_{ref}$ = 2.0 (solid circles); 1.50 (solid squares); 1.25 (upward triangles) and 1.0 (downward triangles).
$\varphi_m$ = $\varphi(\gD = 0)$. In the present work $\varphi_m\;\simeq$  0.585.
Inset: reduced critical volumic fraction defined as $\chi(\varphi,\ep_d)/\chi(\ep_d=0)$ = 0.80
in terms of the median particle size. 
\item[Figure \ref{m_diam_0.05}] 
Magnetization in terms of the applied field for different values of the median diameter,
$\gD/d_{ref}$ = 0.20, $\beta^*$ = 1.0 and $N_p$ = 1007 in the quasi monodisperse case, $\s$ = 0.05.
The corresponding non interacting curves (diameter distribution weighted Langevin curves)
for $d_m/d_{ref}$ = 1.0 (long dash), 1.33 (short dash) and 2.0 (dotted line)
are displayed for comparison.
\item[Figure \ref{m_sigma_1.33}] 
Magnetization in terms of the applied field for different values of the $ln(d)$ standard
deviation $\s$ and $d_m/d_{ref} = 1.33$, $\gD/d_{ref}$ = 0.20 and $\beta^*$ = 1.0.
$N_p$ = 1007 ($\s$ = 0.05), 923 ($\s$ = 0.28); 985 ($\s$ = 0.40) and 990 ($\s$ = 0.50).
The dotted lines are quides to the eye.
The solid lines are the asymptotic limits for $\s$ = 0.05 (bottom) and $\s$ = 0.50 (top).
\item[Figure \ref{m_sigma_1.00} ]
Magnetization in terms of the applied field for
$\s$ = 0.05, 0.20, 0.28 and 0.40 from bottom to top.
and $d_m/d_{ref} = 1.00$, $\gD/d_{ref}$ = 0.20, $\beta^*$ = 1.0.
The dotted lines are guides to the eyes and the solid lines are the corresponding asymptotic
limits for $\s = 0.05$ and $\s = 0.40$.
\item[Figure \ref{anis_0.05_1.0-1.33}] 
Magnetization curve with the anisotropy
energy $\ep_K^{(0)}$ = 2.38, $\s$ = 0.05 and the particles sizes as indicated. 
The solid lines correspond to $\ep_K$ = 0. $\beta^*$ = 1 and $N_p$ = 1007.
\item[Figure \ref{anis_1.33_0.05-0.35}] 
Magnetization curve with the anisotropy
energy $\ep_K^{(0)}$ = 2.38, $d_m/d_{ref}$ = 1.33 and $\beta^*$ = 1 in the polydisperse case 
with $\s$ = 0.35 (open circles); 0.28 (triangles); 0.20 (squares)
compared to the quasi monodisperse case, $\s$ = 0.05 (solid line).
Insert: detail of the downward magnetization curve  in the vicinity of $h$ = 0, showing 
the evolution of the remanence and
coercivity with $\s$ for $\s$ = 0.35, 0.28, 0.20 and 0.10 (open squares) from top to bottom.
\item[Figure \ref{m-h_exp_lf}] 
Experimental low field magnetization curves at room temperature for powder samples of 
$\gamma-$Fe$_2$O$_3$ from ref. \cite{de-montferrand_2012} with polydispersity 
$\s\;\simeq$ 0.26 and median sizes as indicated.
\item[Figure \ref{comp_exp_d10}] 
Comparison between MC simulations and the experimental magnetization curve at 
room temperature for $d_m$ = 10 $nm$. The simulation are performed with $\beta^*$ = 1, 
$\ep_K^{(0)}$ = 2.38 and $N_p$ = 923.
\item[Figure \ref{comp_exp_d12}] 
Comparison between MC simulations and the experimental magnetization curve at 
room temperature for $d_m$ = 12 $nm$. The simulations are performed with $\beta^*$ = 1, 
$N_p$ = 980, $\ep_K^{(0)}$ = 2.38 (open circles) or 1.19 (dashed line).
The dooted line is a guide to the eye.
\item[Figure \ref{comp_exp_d20}] 
Comparison between MC simulations and the experimental magnetization curve at 
room temperature for $d_m$ = 21 $nm$. The simulations are performed with $d_m$ = 20 nm,
$\beta^*$ = 1, $N_p$ = 998, $\ep_K^{(0)}$ = 2.38 (open circles) or 1.19 (dashed line).
\end{itemize} 
\end{singlespace} 
\newpage
\end{document}